\documentclass[aps,prl,showpacs,twocolumn,groupedaddress]{revtex4}

\usepackage{graphicx,bbm}

\def\tr{\,{\rm tr}\,}

\begin{document}

\title{Is efficiency of classical simulations of quantum dynamics related to integrability?}

\author{Toma\v z Prosen and Marko \v Znidari\v c}

\affiliation{
Department of Physics, Faculty of Mathematics and Physics, University of Ljubljana, SI-1000 Ljubljana, Slovenia}

%\author{}
%\email[]
%\thanks{}
%\affiliation{}

\date{\today}

\begin{abstract}

Efficiency of time-evolution of quantum observables, and thermal states of quenched hamiltonians, is studied using {\em time-dependent density matrix renormalization group} method in a family of generic quantum spin chains which undergo a transition from integrable to non-integrable - quantum chaotic case as control parameters are varied.  Quantum states (observables) are represented in terms of matrix-product-operators with rank $D_\epsilon(t)$, such that evolution of a long chain is accurate within fidelity error $\epsilon$ up to time $t$. We find that the rank generally increases {\em exponentially} $D_\epsilon(t) \propto \exp({\rm const}\,t)$, unless the system is integrable in which case we find {\em polynomial} increase.

\end{abstract}

%\keywords{}

\maketitle

In the theory of classical dynamical systems there is a fundamental difference between integrable and chaotic systems. Chaotic systems, having positive algorithmic complexity, unlike the 
integrable ones, cannot be simulated for arbitrary times with a finite amount of information about 
their initial states. Computational complexity of individual chaotic trajectories is linear in time, however, if one wants to describe statistical states (phase space distributions) or observables of chaotic classical systems, up to time $t$, exponential amount of computational resources $N(t) \sim \exp(h t)$ is needed, where $h$ is the Kolmogorov's dynamical entropy related to exponential sensitivity to initial conditions. For example, one needs to expand the solution of the Liouville equation into the lowest $N(t)$ Fourier modes.

How difficult is it to simulate isolated and bounded quantum systems of many interacting particles 
using classical resources? 
%We mean large quantum systems composed of many (or infinite number of)
% interacting particles, since in the opposite case quantum time evolution is almost periodic due to
% discrete energy spectrum. 
In analogy with the classical (chaotic) case, we might expect that the best classical
simulation of typical quantum systems (in thermodynamic limit (TL)) is exponentially hard.
Even though there is no exponential sensitivity to initial conditions in quantum mechanics,
there is a tensor-product structure of the many-body quantum state space which makes
its dimension to scale exponentially with the number of particles, as opposed to linear scaling in the
classical case. Furthermore, due to intricate quantum correlations (entanglement) generic quantum time evolution cannot be reduced to (efficient) classical computation in terms of non-entangled 
(classical like) states. However, it is not known what amount and form of quantum entanglement is needed in order to prevent efficient classical simulation.
% and how fast it is produced by unitary evolutions with various types of Hamiltonians.

Recently, a family of numerical methods for the simulation of interacting many-body systems
has been developed \cite{MPS} which is usually referred to as
time-dependent density-matrix-renormalization group (t-DMRG), and which has been shown to
often provide an efficient classical simulation of certain interacting quantum systems. 
Simulations of locally interacting one-dimensional quantum lattices were actually shown rigorously to be efficient in the number $n$ of particles \cite{Osborn} (i.e., computation time and memory resources scale as polynomial functions of $n$ at fixed $t$, or up to 
 $t=\mathcal{O}(\log n)$), whereas the scaling of computation time and memory with physical time
 $t$ (in TL $n=\infty$), later on referred to as {\em time efficiency}, has not been
 systematically studied. t-DMRG was shown to be time efficient \cite{errors} only in rather special cases of exactly solvable dynamics (generated with XY spin chain Hamiltonian) and/or for particular choices of initial states, lying either in low-energy-sectors or  in low dimensional invariant subspaces. However, for applications in non-equilibrium statistical mechanics and condensed matter theory, e.g. in transport phenomena, it is of primary importance to understand long-time dynamics of generic interacting quantum systems \cite{Sachdev:04}.

In this Letter we address the question of time efficiency implementing up-to-date version of t-DMRG for a family of Ising spin-1/2 chains in arbitrary oriented magnetic field, which undergoes a transition from integrable (transverse Ising) to non-integrable quantum chaotic regime as the magnetic
field is varied. We focus on evolution of density operators of {\em mixed states}, starting from a thermal state of a quenched hamiltonian, and evolution of {\em local} or {\em extensive} initial observables in Heisenberg picture. Note that time evolution of pure states is often ill defined in TL 
\cite{schumacherwerner}. As a quantitative measure of time efficiency we define and compute the minimal dimension 
$D_\epsilon(t)$ of matrix product operator (MPO) representation of quantum states/observables which describes time evolution up to time $t$ within fidelity $1-\mathcal{O}(\epsilon)$. Our central result states that in generic non-integrable
cases computation resources grow exponentially $D_\epsilon(t) \propto \exp(h_{\rm q} t)$, except in the integrable case of transverse Ising chain, where the growth is typically linear $D_\epsilon(t) \propto t$.
 Constant $h_{\rm q}$,  asymptotically independent of $n$, depends only on the evolution (hamiltonian) and not on 
 the details of the initial state/observable or error measures, and can be interpreted as a kind of {\em quantum dynamical entropy}. We conjecture that integrability (solvability) of 1d interacting quantum systems is in one-to-one correspondence with the efficiency of their classical simulability. 

We also studied time efficiency of simulation of {\em pure states} in 
Schr\" odinger picture, for which many examples of efficient applications exist, however all for initial states of rather particular structure typically corresponding to low energy sectors. Treating other, typical states, e.g. eigenstates of unrelated Hamiltonians, linear combinations of highly excited states,  or states chosen randomly in the many-particle
Hilbert space, we found that, irrespectively of integrability of dynamics,
 t-DMRG is {\em not} time-efficient, i.e. $D_\epsilon(t)$ grows exponentially even in the integrable case of transverse field (consistently with a linear growth of entanglement entropy~\cite{Calabrese:05}). 
In view of this fact, our finding that t-DMRG can be time-efficient for integrable systems when implemented for time-evolved operators or high-temperature thermal states, provides a new paradigm fora  successful application of t-DMRG.
\par
Let us briefly review t-DMRG for evolution of density matrices and
operators \cite{tdmrg} which generalizes t-DMRG for pure states
\cite{MPS}. One defines a superket corresponding to an
operator $O$ expanded in the {\em computational basis} of products
of local operators. Concretely, for a chain of $n$ qubits we use a
basis of $4^n$ Pauli operators 
$\sigma^{s_0} \otimes \cdots \otimes \sigma^{s_{n-1}}$, with 
$s_j\in \{0,{\rm x,y,z}\}$ and $\sigma^0 = \mathbbm{1}$. 
The key idea of t-DMRG is to represent any
operator in a matrix product form, 
$O = \sum_{s_j} \tr{(A^{s_0}_0\cdots A^{s_{n-1}}_{n-1})}\,
\sigma^{s_0} \otimes \cdots \otimes \sigma^{s_{n-1}},$ 
in terms of $4 n$ matrices $A^{s_j}_j$ of
fixed dimension $D$. The number of parameters in the MPO representation
is $4nD^2$ and for sufficiently large $D$ it can
describe any operator. In fact, the minimal $D$ required equals to the
maximal rank of the reduced super-density-matrix over all bi-partitions
of the chain. The advantage of MPO representation lies in the fact that
doing an elementary local one or two qubit unitary transformation 
$O'=U^\dagger O U $ can be done locally, affecting only a pair of neighboring
matrices $A^{s_j}_j$.
\par
\begin{figure}[h!]
\includegraphics[angle=-90,width=1.0\linewidth]{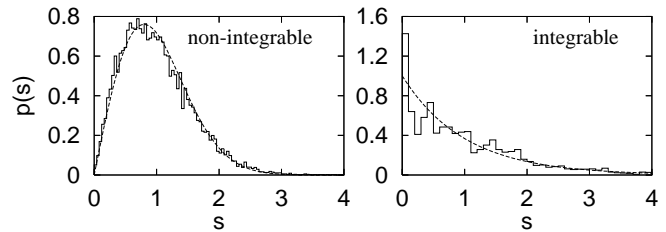}\hfill
\caption{Nearest neighbor LSD for $H_{\rm C}$ (left) and $H_{\rm R}$
(right) for $n=12$. Dashed curves are $p(s)=s \pi/2 \exp{(-\pi^2
s^2/4)}$ (left) and $p(s)=\exp{(-s)}$ (right), typical for chaotic and
regular systems, respectively\cite{Haake:00}. Eigenenergies $\in
[-9,9]$ were used and statistics for even and odd parity states were
combined.} \label{fig:lsd} \end{figure}

 In order to study the role of integrability on
the efficiency of t-DMRG we take antiferromagnetic Ising chain in a
general homogeneous magnetic field, \begin{equation} H(h^{\rm x},h^{\rm
z})=\sum_{j=0}^{n-2}{\sigma^{\rm x}_j \sigma^{\rm x}_{j+1}+
\sum_{j=0}^{n-1} (h^{\rm x}\sigma^{\rm x}_j+h^{\rm z}\sigma_j^{\rm
z}}), \label{eq:H} \end{equation} where $\sigma^{s}_j
=\mathbbm{1}^{\otimes j} \otimes \sigma^s \otimes \mathbbm{1}^{\otimes
(n-1-j)}$. We will analyze evolution for two different magnetic field
values: (i) an integrable (regular) case $H_{\rm R}=H(0,2)$ with
transverse magnetic field and (ii) non-integrable (quantum chaotic)
case $H_{\rm C}=H(1,1)$ with tilted magnetic field. Particular value of
$h^{\rm z}=2$ in the case of $H_{\rm R}$ plays no role. $H_{\rm R}$ can
be solved by Jordan-Wigner transformation which maps $H_{\rm R}$ to a
system of noninteracting fermions. To confirm that $H_{\rm C}$, and
$H_{\rm R}$, indeed represent generic quantum chaotic, and regular,
system, respectively, we calculated level spacing distribution (LSD) of
their spectra (shown in Fig.~\ref{fig:lsd}). LSD is a standard
indicator of quantum chaos~\cite{Haake:00}. It displays characteristic
level repulsion for strongly non-integrable quantum systems, whereas
for integrable systems there is no repulsion due to existence of
conservation laws and quantum numbers.  \par Evolution by t-DMRG
proceeds by splitting hamiltonian (\ref{eq:H}) into even and odd terms,
$H=H_{\rm e}+H_{\rm o}$, such that terms within $H_{\rm e}$ or $H_{\rm
o}$ commute between each other. An approximate propagator for short
time-step is then written using Trotter-Suzuki formula as $U(\delta
t)={\rm e}^{-{\rm i} H_{\rm e} \delta t/2} {\rm e}^{-{\rm i}H_{\rm
o}\delta t} {\rm e}^{-{\rm i}H_{\rm e}\delta t/2}$, where each of the
three terms can be written as a series of commuting one and two qubit
operations. There are two sources of errors in t-DMRG scheme. One is
Trotter error scaling as $\propto (\delta t)^3$ per time step, or
$\propto (\delta t)^2$ in total, and the other, usually dominating one,
is due to truncation. 
\par
\begin{figure}[h!]
\includegraphics[width=0.7\linewidth,angle=-90]{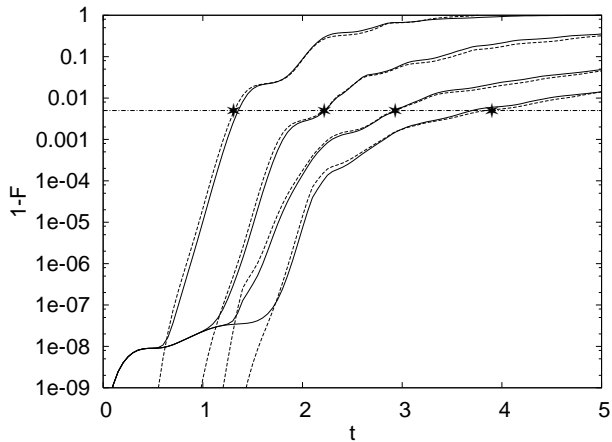}\hfill
\caption{Fidelity (\ref{eq:F}) of t-DMRG evolution (full curves) and
scaled truncation errors $c \eta_{\rm tot}(t)/\delta t$ with $c=0.5$
(dashed curves), for $O(0) = \sigma^{\rm y}_{n/2}$, hamiltonian $H_{\rm
R}$ and $n=10$. Different sets of curves are for $D=10,20,30,40$ (top
to bottom). Chain line marks the threshold where the truncation error
$\eta_{\rm tot}(t)=10^{-4}$ (indicated by stars for different $D$'s).}
\label{fig:testFid} \end{figure}

 The truncation error arrises
because after performing two qubit transformation on MPO the required
dimension of the new matrices increases to $4D$. In order to prevent
the exponential growth of $D$ with time we truncate the resulting
matrices back to dimension $D$~\cite{MPS}. Truncation after application
of a single gate $U_i$ introduces a norm error $\eta(U_i)$ equal to the
sum of squares of discarded singular values. As an estimate for the
total truncation error $\eta_{\rm tot}(t)$ at time $t$ we will use a
sum of all truncation errors $\eta(U_i)$ for two qubit gates $U_i$
applied upto time $t$, $U(t)=\prod_i U_i$ (the number of such gates
scales as $\sim t/\delta t$). If $\lambda^2_k(U_i),
k=0,\ldots 4D-1,$ denote decreasingly ordered eigenvalues of the reduced
super-density-matrix after the application of a gate $U_i$, then
$\eta(U_i)$ and $\eta_{\rm tot}$ are given by \begin{equation}
\eta(U_i) = \sum_{j=D}^{4D-1} \lambda_j^2(U_i),\qquad \eta_{\rm
tot}(t)=\sum_i \eta(U_i). \label{eq:eta} \end{equation} Simple
perturbation argument shows that for small time step $\delta t$, single
gate truncation error scale as  $\eta(U_i) \propto (\delta t)^2$, so
the total error $\eta_{\rm tot}(t) \propto \delta t$. We use the same
time step $\delta t=0.01$ in all our simulations. One may hope that
$\eta_{\rm tot}(t)$ gives a good measure of fidelity \begin{equation}
F(t)=\frac{|\tr{\{O_{\rm MPO}(t) O_{\rm
exact}(t)\}}|^2}{|\tr{\{O^2_{\rm MPO}(t)\}}||\tr{\{O^2_{\rm
exact}(t)\}}|}, \label{eq:F} \end{equation} where $O_{\rm MPO}(t)$ is
an operator obtained from the initial $O$ with t-DMRG evolution with a
given fixed $D$, while $O_{\rm exact}(t)=U^\dagger(t)O U(t)$ is
obtained with an exact evolution. Indeed, by comparing to exact
numerical simulations of small systems of size $n=6,8,10$ and several
different $D$ we find quite generally (see Fig.~\ref{fig:testFid} for
an example) that up to good numerical approximation $1-F(t) \approx
c\eta_{\rm tot}(t)/\delta t$, where $c$ is some numerical constant of
order $1$ which does not depend on $\delta t$, $D$ or $n$.

The central quantity we are going to study is $D_\epsilon (t)$ which is the minimal dimension $D$ of matrices $A^{s_i}_i$ in order for the total truncation error $\eta_{\rm tot}(t)$ to be less than some error tolerance $\epsilon$,
or fidelity (\ref{eq:F}) to be bigger than $1-(c/\delta t) \epsilon $, for  evolution to time $t$. We use $\epsilon=10^{-4}$ for local and extensive operators and $\epsilon=10^{-6}$ for thermal states. 
%Simulations for other values of $\epsilon$ showed equivalent results. 
The central question is: does $D_\epsilon(t)$ grow exponentially or polynomially with $t$? If it grows polynomially we can say that t-DMRG is time efficient.
% As we will see, for integrable $H_{\rm R}$ the growth will usually be polynomial (linear) whereas for %chaotic $H_{\rm C}$ it will be exponential. 
%All our numerical computations are performed on finite chains of lengths begtween 20 and 64, however it has been checked that all results of $D_\epsilon(n)$ are already saturated with increasing $n$ so they in fact probe the thermodynamic limit.

\begin{figure}[h!]
\includegraphics[width=0.7\linewidth,angle=-90]{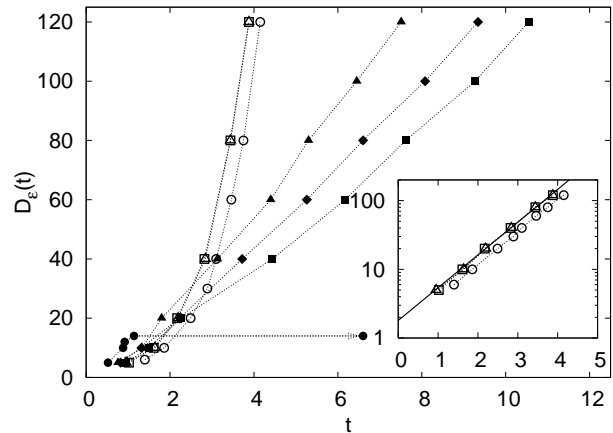}\hfill
  \caption{$D_\epsilon(t)$ for local initial operators.
  We consider three cases $O(0)=\sigma^{\rm x,y,z}_{n/2}$ (empty circles, squares and triangles), for non-integrable evolution $H_{\rm C}$,  and four cases, $O(0)=\sigma^{\rm x,y}_{n/2}$ (full squares, diamonds), $\sigma^{\rm z}_{n/2-1}\sigma^{\rm y}_{n/2}$ (full triangles) with infinite index, 
  and $O(0) = \sigma^{\rm z}_{n/2-1}\sigma^{\rm z}_{n/2}$ (full circles) with index 2, for integrable evolution $H_{\rm R}$. Full line in the inset illustrates exponential growth $\propto e^{1.1 t}$  in the non-integrable case. Full squares and diamonds are for $n=40$, otherwise $n=20$.
%Results shown are for $n=20$ except for $\sigma^{\rm x,y}_{n/2}$ and $H_{\rm R}$ where $n=40$.
%  are practically insensitive to increase of $n$.
  }
\label{fig:local}
\end{figure}

Let us first study the case where the initial operator is a local operator in the center of the lattice
$O(0) = \sigma^s_{n/2}, s\in\{\rm x,y,z\}$. In the integrable case time evolution $O(t)$ can be computed 
exactly in terms of Jordan-Wigner transformation and Toeplitz determinants \cite{jacoby}, 
however for initial operators with infinite {\em index} \cite{explainindex}, 
like e.g for $\sigma^{\rm x,y}_{n/2}$,  $n\to\infty$, the evolution is rather complex and the effective number of terms (Pauli group elements) needed to span $O(t)$ grows exponentially in $t$.
In spite of that, our numerical simulations shown in Fig.~\ref{fig:local} strongly suggest the
linear growth $D_\epsilon(t) \sim t$ for initial operators with infinite index.
Quite interestingly, for initial operators with {\em finite} index, $D_{\epsilon}(t)$ saturates to
a finite value, for example\cite{conjecture} $D_{\epsilon}(\infty)=4$ for $\sigma_{n/2}^{\rm z}$,  or
$D_{\epsilon}(\infty)=16$ for $\sigma_{n/2-1}^{\rm z} \sigma_{n/2}^{\rm z}$.
In non-integrable cases the rank has been found to grow exponentially, $D_\epsilon(t) 
\sim \exp(h_{\rm q} t)$ with exponent $h_{\rm q}$ which does {\em not} depend on
$\epsilon$, properties of $O(0)$ or $n$, for big $n$. For $H=H_{\rm C}$ we find $h_{\rm q} = 1.10$.

\begin{figure}[h!]
\includegraphics[width=0.7\linewidth,angle=-90]{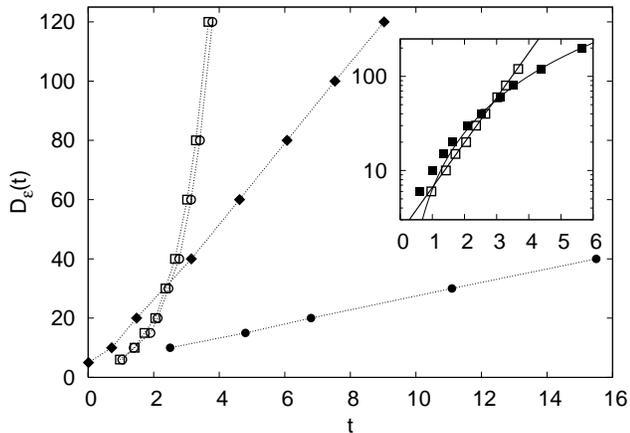}\hfill
  \caption{$D_\epsilon(t)$ for extensive initial operators. For both hamiltonians 
  $H_{\rm C}$, $H_{\rm R}$ we take $O(0)=\sum_j{\sigma^{\rm x}_j}$ (empty, full squares) with
  infinite index, and $O(0)=H(0,1)$ (empty, full circles) with index 1. For $H_{\rm R}$ we also show case $O(0)=\sum_j{\sigma^{\rm z}_j \sigma^{\rm z}_{j+1}+\sigma^{\rm y}_j \sigma^{\rm y}_{j+1}}$ (full diamonds) with index 1 and 2.  In the semi-log inset we illustrate exponential increase $\propto e^{1.1 t}$ (full straight line) for $H_{\rm C}$ and polynomial $\sim t^2$ (full curve) for $H_{\rm R}$. For full circles $n=64$, otherwise $n=32$.}
\label{fig:transl}
\end{figure}

In physics it is often useful to consider extensive observables, for instance
translational sums of local operators, e.g. the hamiltonian $H$ or the total magnetization 
$M^s=\sum_{j=0}^{n-1} \sigma_j^s$. As opposed to local operators, extensive initial operators,
interpreted as W-like super-states, contain some long-range entanglement
so one may expect that t-DMRG should be somewhat less efficient than for local operators.
Indeed,  in the integrable case we find for extensive operators with 
{\em finite} index that  $D_\epsilon(t)$ 
does no longer saturate but now grows linearly, 
$D_\epsilon(t) \sim t$, whereas for extensive operators with {\em infinite} index
the growth may be even somewhat faster, most likely quadratic
$D_\epsilon(t) \sim t^2$ but clearly slower than exponential. 
In the non-integrable case, we again find exponential growth
$D_\epsilon(t) \sim \exp(h_{\rm q} t)$ with the same exponent $h_{\rm q}$ 
as for local initial observables.
The results are summarized in Fig.~\ref{fig:transl}. Note that for local as well as for extensive observables $\eta_{\rm tot}(t)$ asymptotically does not depend on $n$. Therefore the results shown 
in Figs.~\ref{fig:local},\ref{fig:transl}, for which convergence with $n$ has been reached, are already representative of TL. 

\begin{figure}[ht]
\includegraphics[width=0.7\linewidth,angle=-90]{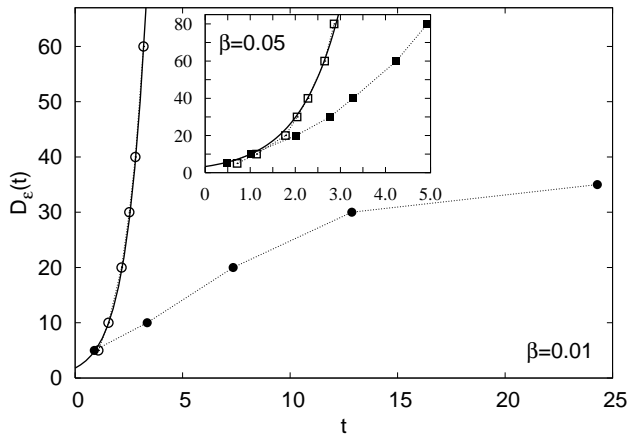}\hfill
  \caption{
  $D_\epsilon(t)$ for thermal states of $H_0$ with $\beta=0.01$ ($\beta=0.05$ in inset),
  for evolution with $H_{\rm C}$ (open symbols) and $H_{\rm R}$ (full symbols) at $n=40$.
  Solid curves again indicate exponential increase $\propto e^{1.1 t}$.}
\label{fig:rhoT}
\end{figure}
In the last set of numerical experiments we consider time efficiency of the evolution of a thermal initial state
$O(t) = Z^{-1} \exp(-\beta H_0)$ under a sudden change of the hamiltonian at $t=0$,
namely $H(t < 0) = H_0=H(0,1), H(t > 0) = H_1$. Again, we treat two situations: in the first case we consider change after which the hamiltonian remains integrable, $H_1 = H(0,2) = H_{\rm R}$, while in the other case the
change breaks integrability of the hamiltonian, $H_1 = H(1,1) = H_{\rm C}$.
Initial state is prepared from identity super-state using imaginary time
t-DMRG with the same MPO rank $D$ as it is later
used for real time dynamics.
We find, consistently with previous results, that at high temperature ($\beta \ll 1$) the
rank $D_\epsilon(t)$ grows very slowly,
perhaps slower than linear, in the integrable case, 
and exponentially $D_\epsilon(t) \sim \exp(h_{\rm q} t)$, in the non-integrable
case. Interestingly, at lower temperatures we find exponential growth in both cases, even in the integrable one. This is not unreasonable as the initial (thermal) state can be
expanded in a power series in $\beta$ and the higher orders $H_0^p$ become less local
with longer entanglement range as we increase the power $p$.
These results are summarized in Fig.~\ref{fig:rhoT}. In contrast to local and W-like observables, the total truncation error $\eta_{\rm tot}(t)$ is for thermal states proportional to $n$. Therefore, the fidelity at fixed $t$ and $D$ of t-DMRG simulation of thermal states decreases in TL.

In conclusion, we have presented numerical experiments suggesting that the scaling of
classical computation resources in t-DMRG simulations of quantum 1d lattices with local interaction
may sensitively depend on the integrability of the hamiltonian, and on whether we propagate pure states 
or mixed states/observables. For the latter we find universal exponential growth of the minimal rank 
of the matrix product representation in physical time, unless we propagate by an integrable hamiltonian from the initial state/observable which can be 
related to (sums of) local operators, in which case the growth is polynomial, or even saturates
for a  specific class of initial operators.
%\begin{acknowledgments}
We acknowledge stimulating discussions with J. Eisert, A.J. Daley, and P. Zoller, and support by Slovenian Research Agency, programme P1-0044, and grant J1-7437.
%\end{acknowledgments}


\begin{thebibliography}{1}

\bibitem{MPS} G.~Vidal, Phys.~Rev.~Lett. {\bf 91}, 147902 (2003); {\em ibid.} {\bf 93}, 040502 (2004); S.~R.~White and A.~E.~Feiguin, Phys.~Rev.~Lett.{\bf 93}, 076401 (2004); A.~J.~Daley, C.~Kollath, U.~Schollw{\" o}ck and G.~Vidal, J.~Stat.~Mech. {\bf 4}, P04005 (2005); 
G.~Vidal, {\tt cond-mat/0605597}.

\bibitem{Osborn} T.~Osborne, {\tt quant-ph/0508031}.

\bibitem{errors}
D.~Gobert, C.~Kollath, U.~Schollw{\" o}ck and G.~Sch{\" u}tz, Phys.~Rev.~E {\bf 71}, 036102 (2005);
U. Schollw{\" ock} and S.~R.~White, in G.~G. Batrouni, and D.~Poilblanc (eds.): Effective models for low-dimensional strongly correlated systems, p.155, AIP, Melville, New York (2006).

\bibitem{Sachdev:04} K.~Sengupta, S.~Powell and S.~Sachdev,
Phys.~Rev.~A {\bf 69}, 053616 (2004).

\bibitem{schumacherwerner} B.~Schumacher and R.~F.~Werner, {\tt quant-ph/0405174}.

\bibitem{Calabrese:05} P.~Calabrese, J.~Cardy, J.~Stat.~Mech. {\bf 04}, P04010 (2005); G.~De Chiara, S.~Montangero,P.~Calabrese and R.~Fazio, J. ~Stat.~Mech.{\bf 5}, P03001 (2006).

\bibitem{tdmrg} F.~Verstraete, J.J.~Garc{\' i}a-Ripoll and J.I.~Cirac, Phys.Rev.Lett. {\bf 93}, 207204 (2004); M.~Zwolak and G.~Vidal, Phys.Rev.Lett. {\bf 93}, 207205 (2004); A.E.~Feiguin and S.R.~White, Phys.Rev.B {\bf 72}, 220401(R) (2005).

\bibitem{Haake:00} F.~Haake, Quantum Signatures of Chaos, Springer Verlag, Berlin (1991) [2nd enlarged Edition 2000].

\bibitem{jacoby} U.~Brandt and K. Jacoby, Z. Phys. B {\bf 25}, 181 (1976); {\em ibid} {\bf 26}, 245 (1977);
J.H.H.~Perk, H.W.~Capel, G.R.W.~Quispel and F.W.~Nijhoff, Physica  {\bf 123A}, 1 (1984).

\bibitem{explainindex} Index of a product operator $O$ [Sect.2, 1st of Refs.\cite{jacoby}] is half the number
of fermi operators in Jordan-Wigner transformation of $O$ and is a conserved quantity for $H_{\rm R}$.

\bibitem{conjecture} Based on numerical results and intuitive arguments we conjecture
that if $O(0)$ is a superposition of finite number of product operators, such that only a single term has the maximal index $\nu$, then the rank saturates at $D_\epsilon(\infty) = 4^\nu$.

\end{thebibliography}
\end{document}